\documentclass[aip,reprint]{revtex4-1}

\newcommand{\blot}[1]{}

\usepackage{float}
\usepackage[section]{placeins}
\usepackage{graphicx}
\usepackage{dcolumn}
\usepackage{bm}

\begin{document}

\preprint{AIP/123-QED}

\title[Gap size and capture zone distributions in one-dimensional point island nucleation and growth simulations]{Gap size and capture zone distributions in one-dimensional point island nucleation and growth simulations: asymptotics and models}

\author{K. P. O'Neill}
 \email{kenneth.o-neill@strath.ac.uk}
 \affiliation{Department of Mathematics and Statistics, University of Strathclyde, Glasgow}
 
\author{M. Grinfeld}
 \email{m.grinfeld@strath.ac.uk}
 \affiliation{Department of Mathematics and Statistics, University of Strathclyde, Glasgow}
 
 \author{W. Lamb}
  \email{w.lamb@strath.ac.uk}
  \affiliation{Department of Mathematics and Statistics, University of Strathclyde, Glasgow}
 
\author{P. A. Mulheran}%
 \email{paul.mulheran@strath.ac.uk}
 \affiliation{Department of Chemical and Process Engineering, University of Strathclyde, Glasgow}


\date{\today}

%


\begin{abstract}
The nucleation and growth of point islands during submonolayer deposition on a one-dimensional substrate is simulated for critical island size $i=0,1,2,3$. The small and large size asymptotics for the gap size and capture zone distributions (GSD and CZD) are studied. Comparisons to theoretical predictions from fragmentation equation analyses are made, along with those from the recently proposed Generalised Wigner Surmise (GWS). We find that the simulation data can be fully understood in the framework provided by the fragmentation equations, whilst highlighting the theoretical areas that require further development. The GWS works well for the small-size CZD behaviour, but completely fails to describe the large-size CZD asymptotics of the one-dimensional system. 
\end{abstract}			      
			      
\maketitle

%

\section{\label{sec_1:intro}INTRODUCTION}

The nucleation and growth of islands during submonolayer deposition
is of considerable theoretical interest as a
fundamental problem in the statistical mechanics of growth processes
\cite{Barabasi95, Venables84, Amar01, PAM08, PAM09}. The sizes and spatial
organisation of the nucleated islands ultimately determine the higher-level
structures such as film and nanostructure array
morphologies \cite{Evans06}. A long-established strategy in the
analysis of the statistical properties is to study capture zones of islands,
since these not only reflect spatial organisation but also
determine growth rates of the islands \cite{PAM95, PAM96,
Bartelt98, PAM00, Evans02, PAM04}. Therefore, the
evolution of capture zones during the deposition process has been a focus 
of many recent theoretical works \cite{PE07, Li10, PE10, Shi09, Oliv11}.

Recently, Pimpinelli and Einstein introduced a new theory for the capture zone distribution (CZD) employing the Generalised Wigner Surmise (GWS) from random matrix theory \cite{PE07}, causing some controversy. Oliveira and Reis \cite{Oliv11} have presented simulation results for islands grown on a two-dimensional substrate with critical island size $i=1$ and $2$, providing some support for the proposed Gaussian tail of the CZD \cite{PE07}. However, Li \emph{et. al.} \cite{Li10} presented an alternative theory which yields a modified form for the large-size CZD behaviour, supported by data for the simulated growth of compact islands with $i=1$. This form seems to agree with that found by Oliveira and Reis, contradicting the GWS \cite{Oliv11}. In other work, Shi \emph{et. al.} \cite{Shi09} studied $i=1$ models in $d=1,2,3,4$ dimensions, finding that the CZD is more sharply peaked and narrower than the GWS suggests. Therefore it is by no means established whether the GWS provides a good theoretical basis for understanding the distribution of capture zones found in island nucleation and growth simulations.

The simplified case of point island nucleation and growth in one dimension has proven to be a good test case for theories. For example, Tokar and Dreyss\'{e} \cite{T&D08} have recently used this model to illustrate their accelerated kinetic Monte Carlo algorithm for diffusion limited kinetics, finding excellent scale invariance in the island size distribution. Blackman and Mulheran \cite{BM96} studied the system with critical island size $i=1$, using a fragmentation equation approach. In this system, we can view the substrate as a string of inter-island gaps, and new island nucleation caused by the deposited monomers as a fragmentation of these gaps. Thus in order to understand the CZD, it is important first to be able to describe the gap size distribution (GSD).

In recent work \cite{GLOM11} we have extended the analysis of the fragmentation equations of
[\onlinecite{BM96}] to the case of general $i = 0, 1, \ldots$. We have been able to derive the small
and large size asymptotics of the GSD, and by assuming random mixing of the gaps caused by
the nucleation process, we have also derived the small size asymptotics for
the CZD for general $i$ and the large size behaviour for $i=0$.

One key feature to emerge from this work is that the asymptotic behaviour of the CZD is again different to that of the GWS \cite{PE07}. It therefore is appropriate to ask what support, further to that in reference [\onlinecite{BM96}], for the fragmentation equation approach is offered by Monte Carlo simulations of the system. Recent work by Gonzalez \emph{et. al.} \cite{GPE11} has revisited the case of $i=1$, developing the original fragmentation equation \cite{BM96} and GWS arguments in response to deviations between prediction and simulation. In this work we will explore simulation results for the one-dimensional (1-D) model with $i=0,1,2,3$, and consider the relative merits of the fragmentation theory \cite{GLOM11} and GWS \cite{PE07} approaches.

The paper is organised as follows. In Section II we summarise the relevant
theoretical results \cite{GLOM11, PE07}. In Section III we describe the
Monte Carlo methods used in our work, both for the full simulation of the
island nucleation and growth processes as well as for nucleation
within single gaps. Simulation results are presented in Section IV
and compared to theoretical predictions, and we finish with a summary and
our conclusions in Section V.

\section{\label{sec_2:mod_meth}THEORY AND PREDICTIONS}

The data from MC simulations can be used as a benchmark against which to
test predictions of theories for the GSD and CZD. In the next two
subsections, we will discuss the predictions of two competing theories,
namely the fragmentation equation approach and the GWS. 

\subsection{Fragmentation Equations}

\begin{figure}[!ht]
 \begin{center}
  \mbox{
        \begin{minipage}{2.5in}
        \scalebox{0.4}{\includegraphics{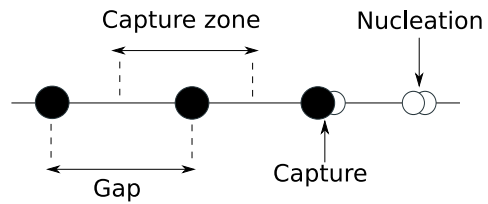}}\\%
	\caption{\label{BM_pic} Summary of the features of the one-dimensional point island model with $i=1$. Solid circles represent an island; open circles are monomers. A capture zone is the separation of the bisectors of neighbouring gaps.\cite{BM96} }
	\end{minipage}
        }\\
 \end{center}
 \end{figure}

We follow the Blackman and Mulheran approach for the 1-D
point-island model with $i=1$ \cite{BM96}. Island nucleation events are
viewed as the fragmentation of gaps between stable islands; see
Figure~\ref{BM_pic}. A nucleation that occurs in a parent gap of width $y$
will result in the creation of two daughter gaps of widths $x$ and $y-x$.
The probability that the nucleation occurs at position $x < y$ is taken from
the long-time (steady state) monomer density profile in the gap,

\begin{equation}
\label{eqBM_mon_profile}
n_1(x)=\frac{1}{2R}x(y-x), 
\end{equation}
where $R=D/F$ is the ratio between the monomer diffusion constant $D$ and the monomer deposition rate $F$. In particular, we assume that the nucleation probability is obtained from this monomer density $n_1(x)^{\alpha}$, with the value of $\alpha$ reflecting the nucleation process. We then obtain \cite{GLOM11}

\begin{eqnarray}
 \label{eqGEE2.1}
 \frac{\partial }{\partial t}u(x,t) & = & -B(\alpha+1,\alpha+1)x^{\lambda}u(x,t) \nonumber \\ 
 & + & 2\int_x^{\infty} [x(y-x)]^{\alpha}u(y,t)dy,
\end{eqnarray}
where $B(\cdot,\cdot)$ is the Beta function and $\lambda=2\alpha+1$. Here, $u(x,t)$ is the number of gaps of size $x$ at time $t$. The first term on the right hand side of (\ref{eqGEE2.1}) is the rate at which gaps of size $x$ are removed from the population by a nucleation event. The second term describes the creation of gaps of size $x$ from the fragmentation of larger gaps, with the factor $2$ reflecting the symmetry of the fragmentation kernel.

In [\onlinecite{GLOM11}] we set $\alpha = i+1$ under the assumption that
nucleation is a rare event solely driven by the diffusion of the monomers.
In doing this we implicitly assume that the $i+1$ monomers necessary to
create the nucleus are all in some sense {\it mature}, each separately
obeying the long-time steady-state density profile $n_1(x)$. However, we
shall also have need to consider the case when nucleation is triggered by a
deposition event. Here a newly deposited monomer either lands close to (or
even directly onto) a pre-existing cluster of $i>0$ mature monomers. In this
case, we set $\alpha=i$. 

Eqn.~(\ref{eqGEE2.1}) admits similarity solutions of the form \cite{ZM86, CR88, EMR05, GLOM11}

\begin{equation}
 \label{eqGEE2.3}
 u(x,t) = \bar{x}(t)^{-2}\phi(x/\bar{x}(t)),
\end{equation}
where $\bar{x}(t)$ is the average gap size. The following asymptotics are
then found \cite{GLOM11}:
 
 \begin{equation}
 \label{eqGSD_small}
  \phi(z) \sim kz^{\alpha} \mbox{ as } z \rightarrow 0;
\end{equation}
\begin{equation}
 \label{eqGSD_large}
 \phi(z) \sim kz^{-2}\exp(-cz^{\lambda}) \mbox{ as } z \rightarrow
 \infty,
 \end{equation}
for constants $c>0$ and $k$. Here $z=x/\bar{x}(t)$ is the scaled gap size.

We may use this information to understand the scaling asymptotics of the CZD
$P(s)$ where $s$ is the scaled capture zone size. On a 1-D
substrate, a point island's capture zone is made up from half of the gap to
its left combined with half of its gap to the right (see Fig.~\ref{BM_pic}). If there
is no correlation between the sizes of two neighbouring gaps, we can write
\cite{BM96}

\begin{equation}
 \label{eqBM}
 P(s) = 2 \int_0^{2s} \phi(z)\phi(2s-z)dz.
\end{equation}
The factor $2$ is included to preserve the normalisation for $P(s)$. 
The small-size asymptotics of $P(s)$ is then \cite{GLOM11}

 \begin{equation}
 \label{eqCZD_small}
  P(s) \sim ks^{2\alpha +1}  \mbox{ as } s \rightarrow 0,
 \end{equation}
for some constant $k$. The large size scaling of $P(s)$ can be computed only for the special case $\alpha=1, i=0$. 
It has been shown that \cite{GLOM11}, for some constant $k$,

 \begin{equation} \\
 \label{eqCZD_large}
  P(s) \sim ks^{-9/2}e^{-2s^3/\mu^3} \ \mbox{ as } s \rightarrow \infty,
 \end{equation}
where $\mu$ is a positive constant.

We note here that the large-size asymptotics of the GSD and the CZD are thus
the same for spontaneous nucleation. Given the form of Eqn.~(\ref{eqBM}) for
$P(s)$, we conjecture that the correspondence between the GSD and CZD
large-size asymptotics will hold for other values of $\alpha=2,3,4...$
although it has not been proved.

\subsection{Generalised Wigner Surmise}

Recently, Pimpinelli and Einstein \cite{PE07} conjectured that the CZD is
well described by the Generalised Wigner Surmise (GWS)
formula, which depends only on one parameter $\beta$
that reflects the critical island size $i$ and the
dimensionality $d$ of the substrate:

\begin{equation}
\label{eqPnE1}
 P_{\beta}(s) = a_{\beta}s^{\beta}\exp(-b_{\beta}s^2),
\end{equation}
where $\beta$ is given by

\begin{equation}
 \label{eqBeta}
 \beta \;
  =
  \left\{
  \begin{array}{cc}
  \frac{2}{d}(i+1) \ \mbox{ if } d = 1,2 \\
  (i+1) \ \mbox{ if } d=3.
  \end{array}
  \right.
\end{equation}

Here $a_{\beta}$ and $b_{\beta}$ are normalisation constants so that 
\[
\int_0^{\infty} P_{\beta}(s)ds = \int_0^{\infty} sP_{\beta}(s) ds = 1.
\]

The remarkable feature of this conjecture is its universal nature; unlike the fragmentation equation approach described above, which is specific to the 1-D substrate, the GWS is claimed to hold for all dimensions. Pimpinelli and Einstein \cite{PE07} demonstrated good agreement with simulation results taken from the literature \cite{BM96, PAM00}, but only with $i=1$ in $d=1$. In their most recent work \cite{GPE11}, this group analyse the i=1, d=1 model in more detail and modify equations (\ref{eqPnE1}) and (\ref{eqBeta}) in response to their findings. Here we note that the asymptotics of the fragmentation equation approach above and the GWS do not agree \cite{GLOM11} for all $i$, whether we adopt $\alpha = i$ or $\alpha = i+1$. This in part motivates the present Monte Carlo study.

\section{\label{sec_3:model_sim}MONTE CARLO SIMULATIONS}

We perform Monte Carlo simulations of point islands on a 1-D
substrate. We adopt the same methodology as in previous work for $i=1$
\cite{BM96}, but here we now also simulate a range of values of the critical
island size $i=0,1,2,3$. In the first subsection we describe the full
simulation for island nucleation and growth, and in the second we describe a
variant for obtaining nucleation rates within a single gap.

\subsection{FULL SIMULATION}

In the full simulation, monomers are randomly deposited at rate $F$
monolayers per unit time onto an initially empty
one-dimensional array of sites representing the substrate. Deposited
monomers diffuse at the rate $D$ by performing random hops between nearest
neighbour lattice sites. We use periodic boundary
conditions. For $i>0$, if the monomer number at any site exceeds the
critical island size, a new island is nucleated. In the case of
spontaneous nucleation ($i=0$), monomers have a small probability
$p_n$ of nucleating a new island each time they hop. Once nucleated, an
island increases in size by absorbing any monomer which hops onto it from a
nearest neighbour site. In the work discussed here,
the islands only ever occupy one lattice site whatever their size in
absorbed monomers. These processes are illustrated in
Fig.~\ref{BM_pic}.

As the deposition rate $F$ increases, the average time a monomer diffuses
before meeting another monomer decreases. Due to the competition between
diffusion and deposition, the statistical properties depend on the ratio
$R=D/F$.

The nominal substrate coverage, $\theta=Ft$, is a useful measure of the
extent of the deposition process. Note that because we simulate point
islands, this coverage can be greater than $100$\% even whilst most of the
substrate remains free for monomer diffusion. For a fixed value of $\theta$,
the average distance between islands increases if $R$ is increased.
Similarly, for fixed $R$, as coverage increases the island density also
increases. We are interested in the scaling properties of the aggregation
regime \cite{Amar94}, where the island density exceeds the monomer density.
The value of $\theta$ for which this regime starts depends on $i$ and $R$,
and we check that the values of $\theta$ employed are sufficiently high to
ensure that we are in the aggregation regime.

Our simulations were performed on lattices with $10^6$ sites, with
$R=8\times 10^6$ up to coverage $\theta=100$\%, averaging results over 100
runs. For $i=0$ we set the spontaneous nucleation probability to
$p_n=10^{-7}$. With these parameters, we find island densities of about $0.5$\%, $1.5$\%, $0.5$\% and $0.25$\% for $i=0,1,2,3$ respectively at $\theta = 100$\%. We note that this is a long way short of the limit referred to by Ratsch \emph{et. al.} \cite{Ratsch05} where scaling breaks down as the lattice becomes saturated with islands. We also have no finite size effects with this size of lattice, and do not need to implement accelerated algorithms \cite{T&D08}.

\subsection{SINGLE-GAP NUCLEATION RATE SIMULATION}

In the single-gap simulation, we simulate island nucleation events in gaps
ranging from size $g=50$ to $g=500$, which proves to be adequate to illustrate the nucleation mechanisms at play. In this variant, monomers
can diffuse as usual on a lattice of length $g$, but are removed from the
simulation if they try to hop beyond the ends of the lattice.

We set the nominal monolayer deposition rate $F$ to unity, so that a monomer
deposition increments the simulated time by $1/g$ (recall that it is the
ratio $R=D/F$ which is important, rather than the absolute value of either
$F$ or $D$). Upon each step of the algorithm, we either deposit a new monomer
at a randomly chosen site in the gap, or diffuse an existing monomer
according to the relative rates of these processes.
Explicitly, a monomer is deposited into
the gap with probability $F\times g /(F\times g+D\times n)=1/(1+R\times
n/g)$, where $n$ is the number of monomers currently in the gap. If no
deposition occurs, a randomly chosen monomer hops to a nearest neighbour
site. If $i+1$ monomers (for $i=1,2,3$) coincide at a site to form a stable nucleus, 
the simulation ends  and the time to the nucleation event recorded. Repeat runs always start with 
an empty lattice, and are used to obtain reliable statistics on the nucleation times within
each gap size.

We use $R=10^6$ for $i=1,2$, and $R=10^5$ for $i=3$ (due to simulation time
constraints). Finally, we also monitor the average monomer density profile
across the gaps, along with the number of hops each monomer makes in the
single-gap simulations. The latter will indicate whether or not island
nucleation is influenced by deposition events. If nucleation is caused solely by the 
diffusional fluctuations of monomers, then the stable nuclei should only include monomers 
that have taken many hops. If however nucleation closely follows a deposition event, then the 
nuclei will contain monomers that have only made few hops since their deposition.

\section{\label{sec_4:results}RESULTS}

\subsection{SINGLE GAP NUCLEATION RATE}

\begin{figure}[H]
 \begin{center}
  \mbox{
        \begin{minipage}{2.5in}
        \scalebox{0.3}{\includegraphics{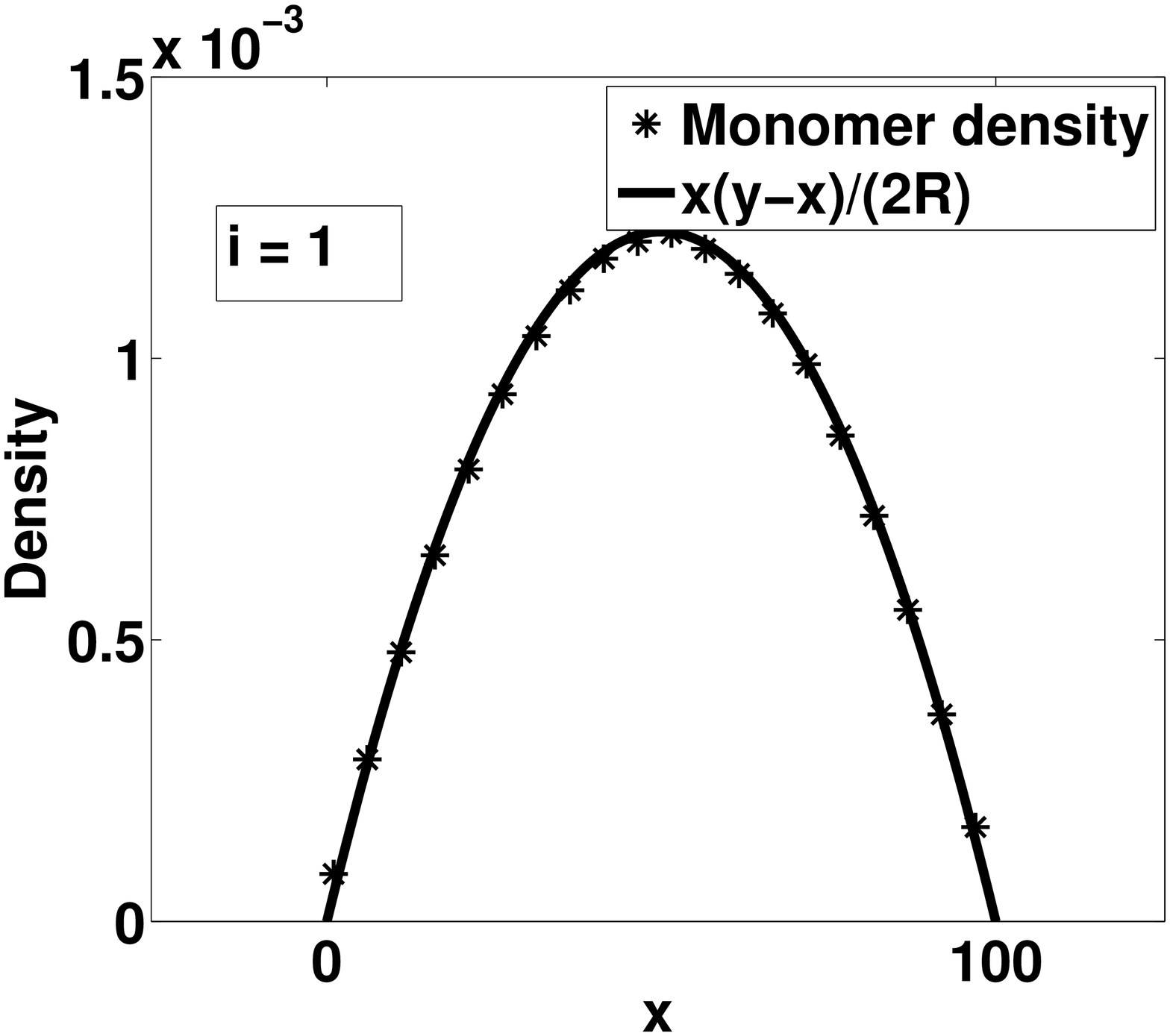}}\\
	\end{minipage}
	}\\
  \end{center}
  \begin{center}
   \mbox{
        \begin{minipage}{2.5in}
        \scalebox{0.3}{\includegraphics{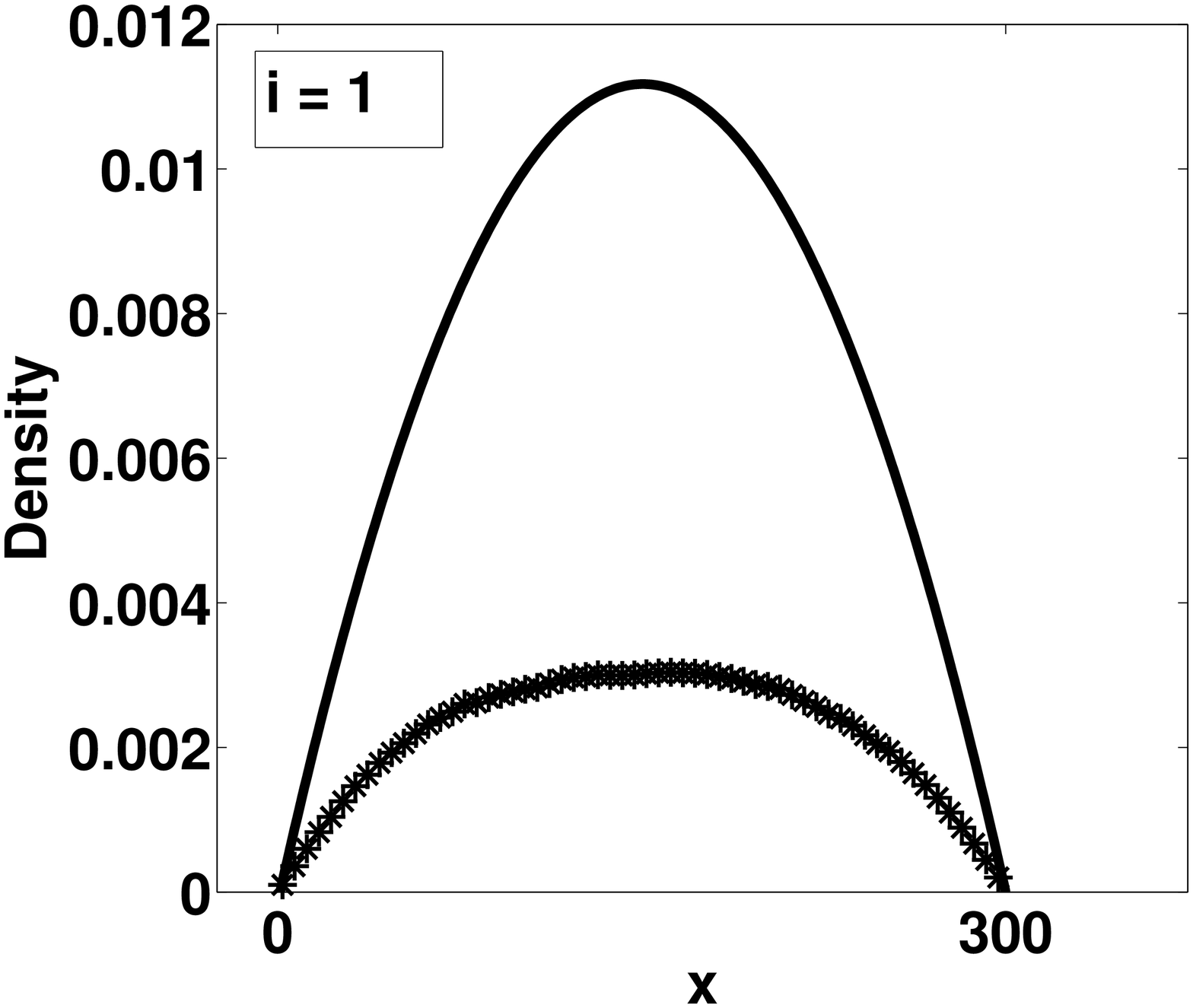}}\\
	\end{minipage}
        }\\
 \end{center}
 \caption{\label{i1d1monomer_density}Monomer density profile in a single gap of size $g=100$ (top) and $g=300$ (bottom) for $i=1$}
\end{figure}

In Fig.~\ref{i1d1monomer_density} we show the results for the average
monomer density profile within gaps of size $g=100$ and $g=300$ for $i=1$. For
the smaller gap size, we see that the profile agrees well with the
assumption made in the fragmentation equation approach \cite{BM96},
coinciding with the long-time steady-state solution of
the diffusion equation with random deposition (Eqn.~(\ref{eqBM_mon_profile})). This is typical for the lower end of the range of
gap sizes that occur in the full simulation at higher coverage, for all the
values of $i$ that we have studied.

However, for the larger gap size $g=300$ shown in
Fig.~\ref{i1d1monomer_density}, we see that the monomer density profile
falls a long way below the long-time prediction. This behaviour is typical
for all values of $i$ at the upper end of gap sizes found in our full
simulations. The reason for the shortfall is the higher nucleation
rate in the larger gaps; the average monomer density profile does
not have sufficient time to reach its saturated level in Eqn.~(\ref{eqBM_mon_profile}) before a nucleation event occurs. As stated, the
range of gap sizes $g$ used in the single-gap simulation is determined by
the range typically seen in our full simulations. Therefore, this failure to
reach the saturated monomer density profile with the large gaps can also be
seen in our full simulation results (data not shown). This will have direct
consequences for how the nucleation rate varies with gap size for larger
gaps, as we now show.

\begin{figure}[H]
 \begin{center}
  \mbox{
        \begin{minipage}{2.5in}
        \scalebox{0.3}{\includegraphics{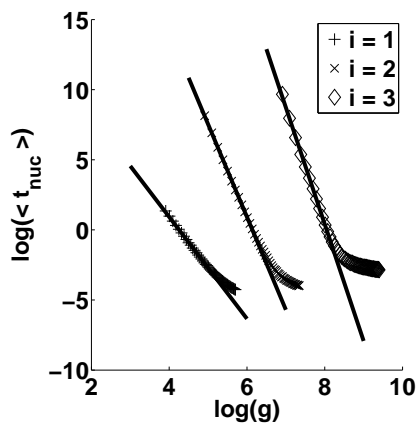}}\\
	\end{minipage}
        }\\
 \end{center}
 \caption{\label{d1nucleation_rate}Average time for a nucleation event to occur at all gaps.}
\end{figure}

In Fig.~\ref{d1nucleation_rate} we show the average time for a nucleation
event to occur $\langle t_{nuc} \rangle$ for all single gaps in the case of
$i=1$, $2$ and $3$ (note that the data for $i=2$ and $i=3$ have been shifted
horizontally to avoid overlapping curves). We note that the data obeys the
power-law form predicted by the fragmentation equation approach for small
gap sizes $g$, but as expected deviates strongly for larger gaps. In fact, the
average time to nucleation becomes much higher than predicted by the use of
the saturated monomer density profile, since the actual profile for the
larger gaps is lower, therefore presenting slower than expected nucleation
rates (but still fast compared to the time it takes for the monomer density
to grow from zero to its saturation level). 

The straight line fits in Fig.~\ref{d1nucleation_rate} are for the small gap
size data only ($g \in [50,150]$).  We use these to estimate how the nucleation
rate varies with gap size $g$ through $1/\langle t_{nuc} \rangle \propto
g^\gamma$, with the values of the power $\gamma$ reported in
Table~\ref{d1nuc_rate}. We have used bootstrap methods with $1000$ samples
of size as big as $80$\% of the original to find an approximate $95\%$
confidence interval in Table~\ref{d1nuc_rate}.

The fragmentation equation approach (Section IIA above) suggests that this
power should be $2i+1$ or $2i+3$, depending on whether island nucleation is
driven by monomer deposition or solely by monomer diffusion. The results in
Table~\ref{d1nuc_rate} suggest that the simulation reflects both these mechanisms, 
with the small gap size nucleation rate exponent lying between these two possibilities. 

\begin{table}[!h]
 \begin{center}
  \begin{tabular}{c|ccc}
   
   $i$ & $\lambda$\footnote{$\lambda=2i+1$} & $\lambda$\footnote{$\lambda=2i+3$} & Simulation \\ \hline
   $1$ & $3$ & $5$ & $3.628 \pm 0.035$ \\ 
   $2$ & $5$ & $7$ & $6.606 \pm 0.118$  \\ 
   $3$ & $7$ & $9$ & $8.283 \pm 0.266$  \\   
  \end{tabular}
  \caption{Small gap nucleation rate exponents from the single gap simulations.}\label{d1nuc_rate}
 \end{center}
\end{table}

\begin{figure}[H]
 \begin{center}
  \mbox{
        \begin{minipage}{2.5in}
        \scalebox{0.32}{\includegraphics{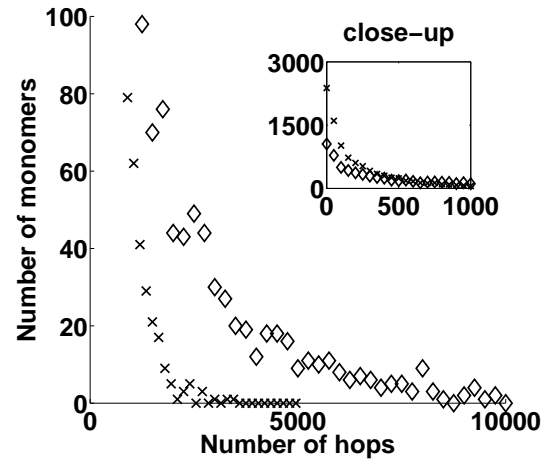}}\\
	\end{minipage}
	}\\
  \end{center}
  \caption{\label{d1no_of_hops}Histogram of the number of hops taken by the youngest monomer in a nucleus for $i=1$, for 
gap size $g=100$ (crosses) and $g=300$ (diamonds). In the main figure the number of monomers is truncated at $100$. The inset shows the same result at the lower number of hops without truncation of the number of monomers in the histogram.} 
\end{figure}

In Fig.~\ref{d1no_of_hops} we present histograms for the number of hops
taken by the youngest monomer in a nucleus for the $g=100$ and $g=300$ $i=1$
simulations. The histogram has a long tail, showing that in many cases all
the monomers in the nucleus are indeed mature in the sense that they have
diffused many times since their deposition. However, there is also a sharp
increase in likelihood of a monomer only taking very few diffusive steps
before being caught up in a nucleation event. In other words, there are a
significant number of nucleation events driven by fluctuations due to
deposition. This supports the conclusion that nucleation in these
simulations is driven by a combination of deposition and diffusion
fluctuations in monomer density, helping to explain the intermediate values for the nucleation
rate exponents in Table~\ref{d1nuc_rate}.

\subsection{FULL SIMULATION BEHAVIOUR}

Having established the nucleation behaviour in single gaps, we can now look at the results observed in our full Monte Carlo simulations. The fragmentation equation approach again provides concrete predictions for the small and large size behaviours for the GSD and CZD. We will also be able to compare the CZD properties with the GWS, and establish which of the two theories provides the better framework to understand the behaviour observed.

\subsubsection{SMALL SIZE SCALING OF THE GSD AND CZD}

\begin{figure}[H]
 \begin{center}
  \mbox{
        \begin{minipage}{2.5in}
        \scalebox{0.3}{\includegraphics{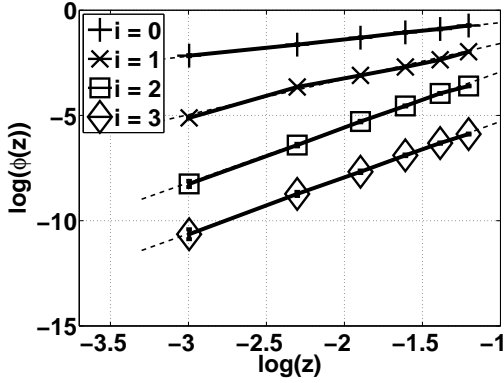}}\\
	\end{minipage}
	}\\
  \end{center}
  \caption{\label{d1small_gsd}Small-size GSD in logarithmic scale for $i=0$, $1$, $2$ and $3$ at coverage $\theta=20$\%. The dashed line is the straight line fit to data.}
\end{figure}

\begin{figure}[H]
 \begin{center}
  \mbox{
        \begin{minipage}{2.5in}
        \scalebox{0.3}{\includegraphics{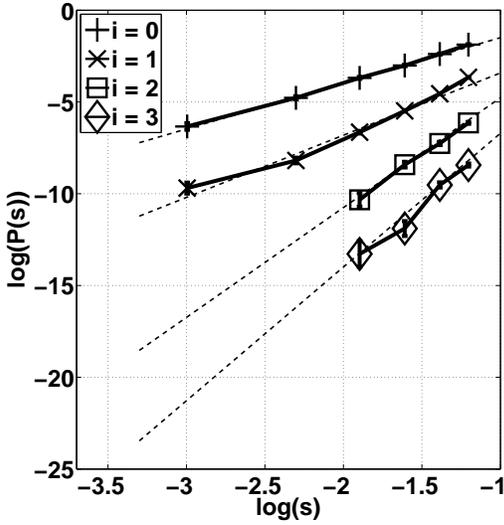}}\\
	\end{minipage}
	}\\
  \end{center}
  \caption{\label{d1small_czd}Small-size CZD in logarithmic scale for $i=0$, $1$, $2$ and $3$ at coverage $\theta=20$\%. The dashed line is the straight line fit to data.}
\end{figure}

In Figures~\ref{d1small_gsd} and \ref{d1small_czd}, we report the small size
behaviour of the GSD ($\phi(z)$) and CZD ($P(s)$) in logarithmic scale at
$\theta=20$\%. In order to fit the slopes in these plots, and obtain
reliable error estimates, we adopt the following numerical technique. The
size data are binned using regularly spaced bins on the logarithmic
abscissa, with bin widths $b^mc$ where $b$ and $c$ are fixed constants and
$m \ge 0$. By choosing a range of values for $b=1.1$, $1.2$, $1.3$ and
$1.4$, and $c=0.0125$, $0.025$ and $0.05$, all of which provide reasonable
choices for binning the data, we obtain a number of straight-line fits. This
allows us to calculate the average of these gradients and a $95$\%
confidence interval. The results of this fitting procedure are shown in
Tables~\ref{d1average_small_GSD_even} and \ref{d1average_small_CZD_even}.

For the small-size asymptotic behaviour of the GSD and CZD we compare the data from MC simulations with the fragmentation equation approach predictions of Section IIA. For the GSD, the dominant term is $z^{\alpha}$ as $z \rightarrow 0$ (see Eqn.~(\ref{eqGSD_small})). Likewise, for the CZD the dominant term is $s^{2\alpha+1}$ (see Eqn.~(\ref{eqCZD_small})). For the latter, we also have the competing prediction of the GWS which is $s^{\beta}$ (Eqn.~(\ref{eqPnE1})). The values from these theories are also displayed in Tables~\ref{d1average_small_GSD_even} and \ref{d1average_small_CZD_even}.

\begin{table}[!h]
 \begin{center}
  \begin{tabular}{c|cccc}
   
   $i$ & $\alpha$\footnote{$\alpha=i$} & $\alpha$\footnote{$\alpha=i+1$} & GSD\footnote{$\theta=20$\%} &  GSD\footnote{$\theta=100$\%}  \\ \hline
   $0$ & - & $1$ & $0.876 \pm 0.033$ & $0.905 \pm 0.029$  \\
   $1$ & $1$ & $2$ & $1.701 \pm 0.045$ & $1.579 \pm 0.105$  \\
   $2$ & $2$ & $3$ & $2.789 \pm 0.080$ & $2.718 \pm 0.074$  \\
   $3$ & $3$ & $4$ & $2.719 \pm 0.082$ & $3.271 \pm 0.056$  \\
  \end{tabular}
  \caption{Average gradient for the small size scaling of the GSD using different bin-widths at coverage $\theta=20$\% and $100$\%}\label{d1average_small_GSD_even}
 \end{center}
\end{table}

\begin{table}[!h]
 \begin{center}
  \begin{tabular}{c|ccccc}
   
   $i$ & $2\alpha+1$\footnote{$\alpha=i$} & $2\alpha+1$\footnote{$\alpha=i+1$} & GWS\footnote{$\beta=2(i+1)$} & CZD\footnote{$\theta=20$\%} &  CZD\footnote{$\theta=100$\%}  \\ \hline
   $0$ & - & $3$ & $2$ & $2.730 \pm 0.030$ & $2.751 \pm 0.086$ \\
   $1$ & $3$ & $5$ & $4$ & $4.187 \pm 0.050$ & $4.372 \pm 0.149$ \\
   $2$ & $5$ & $7$ & $6$ & $5.883 \pm 0.207$ & $5.957 \pm 0.187$ \\
   $3$ & $7$ & $9$ & $8$ & $7.200 \pm 0.382$ & $6.138 \pm 0.124$ \\
  \end{tabular}
  \caption{Average gradient for the small size scaling of the CZD using different bin-widths at coverage $\theta=20$\% and $100$\%}\label{d1average_small_CZD_even}
 \end{center}
\end{table}

The results for the small size scaling exponent of the GSD in
Table~\ref{d1average_small_GSD_even} show that the fragmentation equation
approach provides a reasonably sound framework for understanding the island
nucleation and growth process. For $i=1,2,3$ we see that the exponent at
$\theta=100$\% lies between the two possible values $\alpha=i$ and
$\alpha=i+1$ suggested by the theory. This is as expected following the
single-gap nucleation results presented above, which show that both the
deposition- and diffusion-driven nucleation mechanisms are at play in the
simulations. We note that the $\theta=20$\% results for $i=3$ lie below
$\alpha=i=3$, but we believe that this is due to the fact that the
simulation has only just entered the aggregation regime in this case. We
also see that for $i=0$, the exponent is close to the $\alpha=i+1=1$
prediction ($\alpha=0$ is not a viable possibility), being closer at
$\theta=100$\%.

The trends shown in the small size scaling exponent of the CZD in
Table~\ref{d1average_small_CZD_even} are rather similar. We see the $i=0$
data are close to the $\lambda=2i+3=3$ prediction of the fragmentation
equation approach, being somewhat larger than the $\beta=2(i+1)=2$ predicted
by the GWS. For $i=1,2$ the data are bracketed by the two alternatives
suggested by the fragmentation theory, as indeed is the GWS exponent which
appears to present a reasonable compromise given the two alternative
nucleation mechanisms. The case of $i=3$ provides an exception, which hints
at the breakdown of the relation in Eqn.~(\ref{eqBM}) between the GSD and
the CZD. This will be discussed further in the final section.

\subsubsection{LARGE SIZE SCALING OF THE GSD AND CZD}

\begin{figure}[H]
 \begin{center}
  \mbox{
        \begin{minipage}{2.5in}
        \scalebox{0.3}{\includegraphics{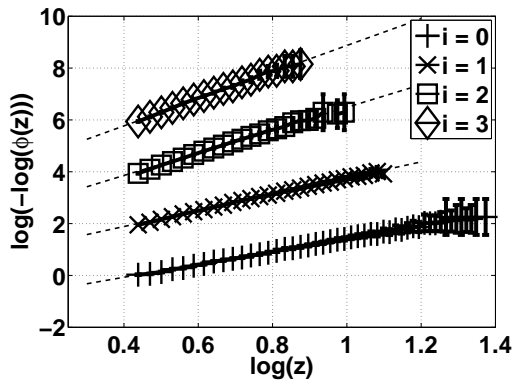}}\\
	\end{minipage}
	}\\
  \end{center}
  \caption{\label{d1large_gsd}Large-size GSD in logarithmic scale for $i=0$, $1$, $2$ and $3$. The dashed line is the straight line fit to data.}
\end{figure}

\begin{figure}[H]
 \begin{center}
  \mbox{
        \begin{minipage}{2.5in}
        \scalebox{0.3}{\includegraphics{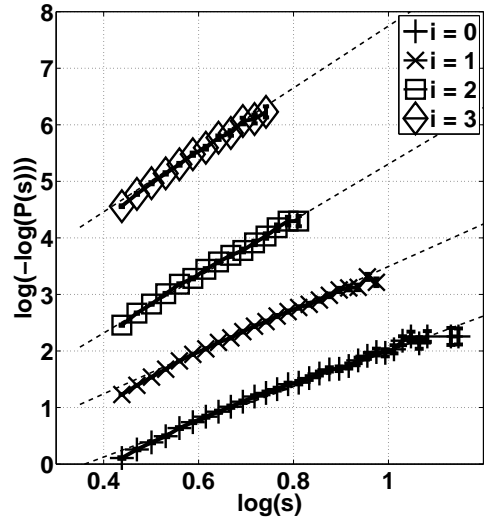}}\\
	\end{minipage}
	}\\
  \end{center}
  \caption{\label{d1large_czd}Large-size CZD in logarithmic scale for $i=0$, $1$, $2$ and $3$. The dashed line is the straight line fit to data.}
\end{figure}

In Figures~\ref{d1large_gsd} and \ref{d1large_czd} we present the large-size behaviour of the GSD and CZD from the full simulations. The data are plotted in order to test the common large-size functional form suggested by the fragmentation equation approach for the GSD and by the GWS for the CZD , namely $\exp(-cz^p)$ (see Eqns.~(\ref{eqGSD_large}) and (\ref{eqPnE1})). In all cases, the data do conform well to this functional form. In addition, we perfom fits to find the gradients $p$ on these plots. In order to provide an estimate of the error in these fits, we adopt a similar strategy to that used above for the small-size scaling and bin the data using binwidths of size $0.01k$ with $k=1,2,...,20$. The results of this fitting procedure are presented in Tables~\ref{d1average_large_GSD} and \ref{d1average_large_CZD} for the GSD and CZD respectively.

Once again we compare the exponents $p$ from the Monte Carlo simulation data with the theoretical predictions. For the GSD, the fragmentation equation approach predicts values of $2\alpha+1$ for $p$. For the CZD, the fragmentation equation prediction is $p=3$ for $i=0$ (Eqn.~(\ref{eqCZD_large})) and we conjecture that the values for $i>0$ will match those of the GSD. In contrast, the GWS prediction for the CZD is the universal value $p=2$. The values from these theories are displayed in Tables~\ref{d1average_large_GSD} and \ref{d1average_large_CZD}. 

\begin{table}[!h]
 \begin{center}
  \begin{tabular}{c|cccc}
   
   $i$ & $2\alpha+1$\footnote{$\alpha=i$} & $2\alpha+1$\footnote{$\alpha=i+1$} & GSD\footnote{$\theta=20$\%} &  GSD\footnote{$\theta=100$\%}  \\ \hline
   $0$ & - & $3$ & $2.515 \pm 0.006$ & $2.665 \pm 0.007$  \\
   $1$ & $3$ & $5$ & $3.130 \pm 0.009$ & $3.383 \pm 0.008$  \\
   $2$ & $5$ & $7$ & $4.364 \pm 0.020$ & $5.112 \pm 0.025$  \\
   $3$ & $7$ & $9$ & $5.094 \pm 0.026$ & $6.437 \pm 0.034$  \\
  \end{tabular}
  \caption{Average exponents for the large size scaling of the GSD using different bin-widths at coverage $\theta=20$\% and $100$\%}\label{d1average_large_GSD}
 \end{center}
\end{table}

\begin{table}[!h]
 \begin{center}
  \begin{tabular}{c|cccc}
   
   $i$ & $2i+3$\footnote{$\lambda=2i+3$} & GWS & CZD\footnote{$\theta=20$\%} &  CZD\footnote{$\theta=100$\%}  \\ \hline
   $0$ & $3$ & $2$ & $3.108 \pm 0.012$ & $3.043 \pm 0.043$ \\
   $1$ & - & $2$ & $3.721 \pm 0.020$ & $3.826 \pm 0.021$ \\
   $2$ & - & $2$ & $4.946 \pm 0.029$ & $5.536 \pm 0.033$ \\
   $3$ & - & $2$ & $5.464 \pm 0.041$ & $6.530 \pm 0.042$ \\
  \end{tabular}
  \caption{Average exponents for the large size scaling of the CZD using different bin-widths at coverage $\theta=20$\% and $100$\%}\label{d1average_large_CZD}
 \end{center}
\end{table}

In Table~\ref{d1average_large_GSD} we see that the fragmentation equation approach provides a useful point of reference to the observed large-size scaling exponents of the GSD. Again we see values that are bracketed by the two possible nucleation mechanisms for $i=1,2$, whilst the behaviour for $i=0$ is a little below the predicted exponent of $p=3$. For $i=3$ the data's exponent is below even that of the deposition-induced nucleation case. However, we have shown in Section~\ref{sec_3:model_sim} above that the monomer density profile does not reach its saturation value in larger gaps, so that the nucleation rate in these gaps is lower than predicted by the theory. This seems to provide a rational explanation for the discrepancies. 

The results in Table~\ref{d1average_large_CZD} for the large-size scaling behaviour of the CZD are rather informative. We firstly observe that the Monte Carlo data exponents do indeed mirror those of the GSD in Table~\ref{d1average_large_GSD} quite well. This means that the universal GWS prediction for $p=2$ is always wrong. We also see that the concrete prediction for $i=0$ from the fragmentation equations, namely $p=3$, is well supported by the simulation data.

\section{\label{sec_5:conclusion}SUMMARY AND CONCLUSIONS}

We have investigated one-dimensional (1-D) point island nucleation and growth simulations in order to test predictions for the asymptotics of the gap and capture zone size distributions (GSD and CZD respectively). The work shows that the fragmentation equation approach provides a good framework in which to understand the Monte Carlo simulation results. The theory can be used to investigate two cases for the nucleation process for $i>0$, the first where nucleation is driven by deposition events, the second where fluctuations caused solely by monomer diffusion induce nucleation.

Firstly we presented single gap simulation results which show that both these nucleation processes are active, so that the observed nucleation rates are bracketed by these two extremes. Furthermore, we showed that for larger gaps the average monomer density profile does not reach the long-time steady state assumed in the fragmentation equations. As a result, the nucleation rates in large gaps are slower than predicted by the theory, with the shortfall increasing with gap size. Therefore, the simple power-law scaling of the nucleation rate with gap size breaks down at larger sizes, with obvious consequences for the fragmentation equation predictions for the GSD.

We note here that deviations from the original Blackman and Mulheran \cite{BM96} predictions for the nucleation rate dependence on gap size have recently been observed for the $i=1$ 1-D model \cite{GPE11}. In this work, the authors report that the nucleation rate has two regimes; for small sizes, it approximately obeys $s^4$, whilst at larger sizes it approximately follows $s^3$. The latter power-law feeds into the asymptotic form of the GSD and hence the CZD, yielding the functional form $\exp(-s^3)$. We note here that these values are close to those we find for $i=1$ in Table~\ref{d1nuc_rate} for the small gap nucleation rates and Tables~\ref{d1average_large_GSD} and \ref{d1average_large_CZD} for the large-size GSD and CZD scaling. We therefore propose that the explanations presented here in terms of competing nucleation mechanisms and unsaturated monomer density profiles will also explain the results reported in [\onlinecite{GPE11}].

We also presented data for the full island nucleation and growth simulation. For the small-size GSD scaling, we found results consistent with the fragmentation equation predictions for $i=0$. For $i=1,2,3$ the exponents were bracketed by the values for the alternative nucleation mechanisms as expected. For the large gap size scaling, the Monte Carlo data followed the functional form suggested by the fragmentation theory, with the exponents again being largely bracketed by the predicted values, although the breakdown of the nucleation rate scaling is apparent, especially for larger $i$. 

In the case of the CZD, we once again successfully placed the observed simulation data into the context provided by the fragmentation equations. Interestingly, the GWS predictions for the small-size CZD scaling work extremely well since they bisect the exponents from the alternative nucleation mechanisms. As discussed elsewhere \cite{GLOM11}, the predicted formula for the parameter $\beta$ of the GWS can be brought into line with either nucleation mechanism following the arguments of Pimpinelli and Einstein \cite{PE10}, but the original prediction of these authors (Eqn.~(\ref{eqBeta})) does seem to speak well for their physical intuition \cite{PE07}.

However, the predicted GWS form for the large-size CZD scaling fails badly when confronted with our 1-D point island simulation results. This is in contrast to recent tests performed using two-dimensional substrates \cite{Oliv11}, which suggests that there is something unique to the 1-D case, possibly due to the topological constraints in how capture zones are constructed from the inter-island gaps. This aspect is worthy of further investigation.

In order to predict the asymptotics of the CZD, we have assumed that the capture zones can be constructed from pairs of gaps sampled randomly for the GSD (see Eqn.~(\ref{eqBM})). This is valid provided that the nucleation has effectively mixed up the gaps so that nearest neighbours are no longer correlated \cite{BM96}. One consequence is that the small-size exponents of the CZD (say $p_1$) are related to those of the GSD (say $p_2$) through $p_1 = 2p_2+1$. Looking at the results in Tables~\ref{d1average_small_GSD_even} and \ref{d1average_small_CZD_even}, we see that this relationship is reasonably obeyed for $i=0,1$ but starts to break down for $i=2,3$. This is perhaps understandable, since for the higher critical island sizes, the nucleation rate slows down dramatically over time suggesting less well-mixed systems. This is another point for further consideration in future theory development work. 

Despite the limitations of the fragmentation equation approach used in this work, such as its failure to capture the time-dependent nature of the monomer density profile within gaps, it has provided an excellent theoretical framework from which to consider the island nucleation process. Hence, alongside the points discussed above, future work might also look at how the fragmentation kernels can incorporate this time dependency, and how the two nucleation mechanisms can be combined into a consistent set of fragmentation equations. 

\begin{acknowledgments}
KPON is supported by the University of Strathclyde through a PhD scholarship. The simulation data were obtained using the Faculty of Engineering High Performance Computer at the University of Strathclyde.
\end{acknowledgments}

http://www.mathstat.strath.ac.uk/

http://www.strath.ac.uk/chemeng/research\\/groupdetails/drpaulmulheran-seniorlecturer/

\bibliography{aipsamp}

\end{document}